# Analytic formula for hidden variable distribution: Complex networks arising from fluctuating random graphs

Sumiyoshi Abe[1,*] and Stefan Thurner[2,**]

[1]*Institute of Physics, University of Tsukuba, Ibaraki 305-8571, Japan*

[2]*Complex Systems Research Group, HNO, Medizinische Universität Wien,*

*Währinger Gürtel 18-20, Vienna A-1090, Austria*

**Abstract**  In analogy to superstatistics, which connects Boltzmann-Gibbs statistical mechanics to its generalizations through temperature fluctuations, complex networks are constructed from the fluctuating Erdös-Rényi random graphs. Here, using the quantum mechanical method, the exact analytic formula is presented for the hidden variable distribution, which describes the fluctuation and generates a generic degree distribution through the Poisson transformation. As an example, a static scale-free network is discussed and the corresponding hidden variable distribution is found to decay as a power law.

In recent years, much attention has been paid to the study of network structures underlying artifactual as well as natural complex systems. There, a vertex and an edge represent an element and a relation between elements, respectively. A main interest here is in the topological structure of the network, and the physical details of interactions between elements play rather marginal roles. This certainly offers an efficient approach to understanding collective behaviors of complex systems. The study along this line was initiated by the works of Watts and Strogatz [1] on small-world networks and of Barabási and Albert [2] on scale-free networks (see also [3]). These models are essentially different from random graphs considered by Erdös and Rényi [4] (see also [5]) least in the following two points: i) a small-world network has a value of the clustering coefficient [1], which is much larger than that of a random graph, showing strong correlation between the edges, and ii) the degree distribution of a scale-free network has the form [2]

$$p_{sf}(k) \sim \frac{1}{(k+k_0)^\gamma} \qquad (1)$$

with vertex connectivity, $k$, and constants, $k_0 \in (0,1)$ and $\gamma > 1$, whereas it is the Poisson distribution for a random graph.

The scale-free degree distribution in Eq. (1) implies that there exist a significant number of vertices that have high values of connectivity (i.e., hubs). Such a structure is profoundly relevant to the concept of robustness and vulnerability of the network [6].

In [2], it has been discussed that it is sufficient for realizing a scale-free network to consider its growth with the so-called preferential attachment rule, that is, a vertex created anew tends to be linked to an old vertex with probability proportional to connectivity of that old vertex. This may be able to explain the structures of the Internet [7], the WWW [3], and pattern of citing papers [8]. It, however, does not seem to explain the natures of biological networks such as food webs [9], metabolic networks [10], and protein architectures in cells [11], for example. This implies that there should be a variety of mechanisms, which can lead to scale-free statistics in Eq. (1).

There are in fact some methods of generating scale-free networks, which do not assume growth and preferential attachment. For example, in a static model given in [12], a weight factor of the specific form is assigned to each vertex. Then, two vertices are randomly selected with the probability proportional to their weights and are connected by an edge if they were not already linked (otherwise, they are discarded). Such an algorithm leads to Eq. (1) with $\gamma > 2$ for large $k$. Another static model proposed in [13] introduces varying vertex fitness. This idea has further been elaborated in [14] (see also [15]). Nowadays, varying fitness is referred to as the hidden variable in the literature.

In this paper, we construct static complex networks by making use of fluctuating random graphs. The discussion may be somewhat similar to the works in [12-15], but it is in the spirit of superstatistics [16-20] that establishes a connection between Boltzmann-Gibbs statistical mechanics and its generalizations based on temperature fluctuations. An essential point here is the analogy of Erdös-Rényi random graph theory to Boltzmann-Gibbs statistical mechanics: in the latter, temperature of a canonical ensemble is fixed, while the probability of connecting each two vertices is fixed in the former. Therefore, temperature fluctuation corresponds to varying probability of connection. Applying the quantum mechanical method, we derive a general formula for the hidden variable distribution, which relates a random graph to a general complex network. As an example, we analyze a scale-free network, in detail, and present the explicit formula for the associated hidden variable distribution. Such a hidden variable distribution is shown to decay as a power law. Thus, the numerical result given in [13] is confirmed analytically.

Let us recall the classical discussion about random graphs of Erdös and Rényi. Consider $N$ vertices. Each pair of vertices is connected by an edge with probability, $q$. New edges are attached in this way, up to the total number, $N-1$. Then, the probability of finding a vertex with degree (i.e., connectivity) $k$ is given by the binomial distribution. In the large $N$ limit, such a degree distribution becomes Poissonian:

$$p_{\text{random}}(k) = \frac{\lambda^k}{k!} e^{-\lambda}, \tag{2}$$

where $\lambda$ is a fixed constant, $q(N-1)$. This is the analog of the canonical distribution in Boltzmann-Gibbs statistical mechanics, and thus $\lambda$ corresponds to temperature.

Now, a question is how the degree distribution can change if $\lambda$ is a stochastic variable, in analogy with temperature fluctuation in superstatistics [16-20]. In this case, the distribution in Eq. (2) is regarded as a conditional probability distribution, $p_{random}(k|\lambda)$. Thus, we have

$$\int_0^\infty d\lambda\, \Pi(\lambda) \frac{\lambda^k}{k!} e^{-\lambda} = p(k), \qquad (3)$$

where $p(k)$ is a certain degree distribution and $\Pi(\lambda)$ is the associated hidden variable distribution [13-15]. In other words, $p(k)$ is the marginal of the joint probability, $\Pi(\lambda) p_{random}(k|\lambda)$. Eq. (3) implies that $p(k)$ is the Poisson transformation of the hidden variable distribution. In this way, we are able to construct a wide class of complex networks from a fluctuating random graph.

First, let us consider a general form of $p(k)$. This is of obvious importance since real-world networks have diverse structures, but they can still be characterized by their degree distributions [21]. To find the form of $\Pi(\lambda)$, we employ the quantum mechanical method. In particular, we use the coherent state of the harmonic oscillator with unit mass and frequency. The Planck constant is also set equal to unity. The creation and annihilation operators, $\hat{a}^\dagger$ and $\hat{a}$, satisfy the algebra: $[\hat{a}, \hat{a}^\dagger] = 1$, $[\hat{a}, \hat{a}] = [\hat{a}^\dagger, \hat{a}^\dagger] = 0$. The ground state, $|0\rangle$, is annihilated by $\hat{a}$: $\hat{a}|0\rangle = 0$. The $k$- particle state is given by $|k\rangle = (k!)^{-1/2}(\hat{a}^\dagger)^k|0\rangle$, which is the eigenstate of the number operator, $\hat{n} = \hat{a}^\dagger \hat{a}$, that is, $\hat{n}|k\rangle = k|k\rangle$. $\{|k\rangle\}_{k=0,1,2\cdots}$ forms an orthonormal complete set termed the Fock basis. The coherent state, $|\alpha\rangle$, is defined by, $|\alpha\rangle = e^{-|\alpha|^2/2}\sum_{k=0}^{\infty}(\alpha^k/\sqrt{k!})|k\rangle$, where $\alpha$ is a complex variable. It also forms the (over)complete set, satisfying $\pi^{-1}\iint d^2\alpha\,|\alpha\rangle\langle\alpha| = 1$, where $d^2\alpha \equiv d(\text{Re}\,\alpha)d(\text{Im}\,\alpha)$ and the domain of integration is the whole complex $\alpha$-plane. It should be noticed that the number distribution in the coherent state is Poissonian: $|\langle k|\alpha\rangle|^2 = \left[\left(|\alpha|^2\right)^k / k!\right]e^{-|\alpha|^2/2}$.

Now, we consider the *P*-representation of a density matrix $\hat{\rho}$ [22,23]:

$$\hat{\rho} = \iint d^2\alpha\, P(\alpha,\alpha^*)|\alpha\rangle\langle\alpha|. \qquad (4)$$

The function, $P(\alpha,\alpha^*)$, may be singular and negative, in general (however, as we shall see, it turns out to behave well as a probability distribution in the present context). Assume that $\hat{\rho}$ is a certain function of the number operator,

$$\hat{\rho} = p(\hat{n}). \qquad (5)$$

Then, from Eq. (4), it follows that

$$\iint d^2\alpha \, P(\alpha, \alpha^*) \frac{\left(|\alpha|^2\right)^k}{k!} e^{-|\alpha|^2/2} = p(k). \tag{6}$$

In this representation, the degree distribution is the analog of the energy (i.e., particle number) distribution. It would be of interest to examine possibility of applying the second quantization formalism to constructing a network based on such an analogy. Since $\hat{\rho}$ is the function of the number operator, $P(\alpha, \alpha^*)$ is a function only of $|\alpha|^2$. Therefore, we see the parallelism between Eqs. (3) and (6) with identification, $\lambda \leftrightarrow |\alpha|^2$. A crucial point is that it is possible to express $P(\alpha, \alpha^*)$ in Eq. (4) conversely in terms of $\hat{\rho}$ as follows [23]:

$$P(\alpha, \alpha^*) = \frac{1}{\pi^2} e^{|\alpha|^2} \iint d^2\beta \, \langle -\beta|\hat{\rho}|\beta\rangle \, e^{|\beta|^2} e^{\alpha\beta^* - \alpha^*\beta}. \tag{7}$$

From this equation, we can arrive at the general analytic formula for the hidden variable distribution

$$\Pi(\lambda) = \pi P(\alpha, \alpha^*)$$
$$= \frac{1}{\pi} e^{|\alpha|^2} \iint d^2\beta \, e^{\alpha\beta^* - \alpha^*\beta} \sum_{k=0}^{\infty} p(k) \frac{\left(-|\beta|^2\right)^k}{k!} \tag{8}$$

with $\lambda \equiv |\alpha|^2$.

Finally, let us calculate as an important example the hidden variable distribution for a scale-free network characterized by the degree distribution in Eq. (1). In this case, the normalized density matrix is taken to be

$$\hat{\rho}_{sf} = \frac{A}{(\hat{n}+k_0)^\gamma},  \quad (9)$$

$$A^{-1} = \zeta(\gamma, k_0),  \quad (10)$$

where $\zeta(s, a)$ is Hurwitz's generalized zeta function [24], which is well defined for $s > 1$ (but can be analytically continued to an arbitrary complex $s$ except for the singularity at $s = 1$). Using the Mellin transformation of $\hat{\rho}_{sf}$, that is, $\hat{\rho}_{sf} = \left[\zeta(\gamma, k_0)\Gamma(\gamma)\right]^{-1} \int_0^\infty dt\, t^{\gamma-1} e^{-(\hat{n}+k_0)t}$ with Euler's gamma function $\Gamma(z)$, we obtain the following analytic formula for the hidden variable distribution:

$$\Pi_{sf}(\lambda) = \frac{1}{\zeta(\gamma, k_0)\Gamma(\gamma)} \int_0^\infty dt\, t^{\gamma-1} e^{(1-k_0)t - (e^t - 1)\lambda}.  \quad (11)$$

The first and second moments of this distribution are analytically calculated to be

$$\langle \lambda \rangle = \frac{\zeta(\gamma-1, k_0)}{\zeta(\gamma, k_0)} - k_0,  \quad (12)$$

$$\langle \lambda^2 \rangle = \frac{\zeta(\gamma-2, k_0)}{\zeta(\gamma, k_0)} - (2k_0+1)\frac{\zeta(\gamma-1, k_0)}{\zeta(\gamma, k_0)} + k_0(k_0+1), \qquad (13)$$

respectively. From Eq. (11) as well as these, it is clear that $\Pi_{sf}(\lambda)$ decay as a power law

$$\Pi_{sf}(\lambda) \sim \frac{1}{\lambda^\delta} \qquad (14)$$

with the exponent

$$\delta = \gamma. \qquad (15)$$

In Fig. 1, we present the plot of $\Pi_{sf}(\lambda)$ with respect to $\lambda$. There, one recognizes the asymptotic power-law behavior of $\Pi_{sf}(\lambda)$. This, in turn, explains why the authors of [13] could numerically realize a scale-free network by using a power-law hidden variable distribution.

In conclusion, we have constructed in the spirit of superstatistics a generic complex network by introducing fluctuation to a random graph. We have derived the exact analytic formula for the hidden variable distribution, which describes the fluctuation and generates a general form of the degree distribution through the Poisson transformation. As an example, we have considered a static scale-free network. We have explicitly calculated the associated hidden variable distribution and have shown that it decays as a power law with the same exponent as that of the degree distribution.

S. A. was supported in part by the Grant-in-Aid for Scientific Research of Japan Society for the Promotion of Science. S. T. acknowledges support from the FWF Grant P17621.

# Figure caption

Fig. 1    The log-log plot of $\Pi_{sf}(\lambda)$ with respect to $\lambda$ with $\delta\,(=\gamma)=2.5$ and $k_0=0.5$. All quantities are dimensionless.

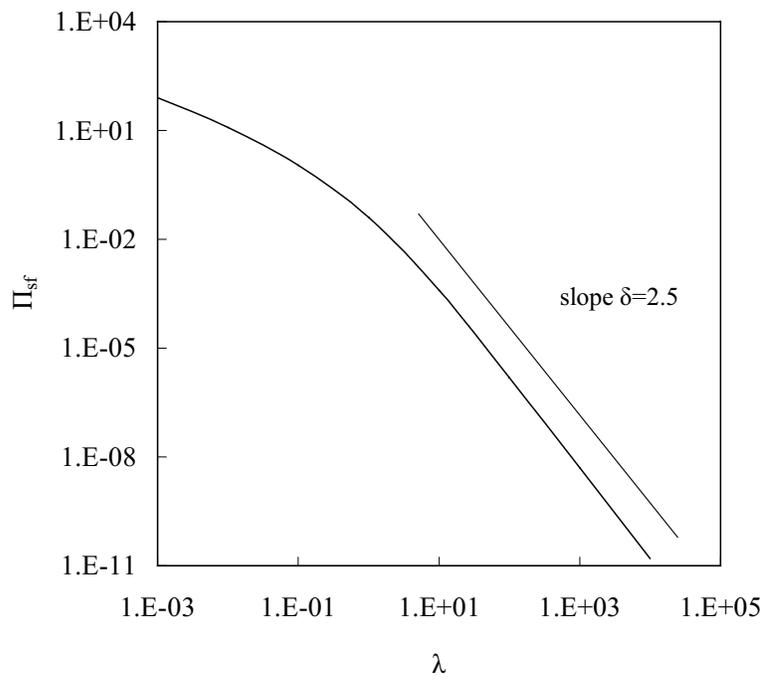

Fig. 1